\documentclass[fleqn,usenatbib]{mnras}
\usepackage[utf8]{inputenc}
\usepackage{newtxtext,newtxmath}
\usepackage[T1]{fontenc}
\usepackage{graphicx}
\usepackage{amsmath}
\usepackage{booktabs}

\begin{document} 
   \title[Characterisation of Timau National Observatory]{Characterisation of Timau National Observatory using limited \emph{in-situ} measurements}
   \author[R. Priyatikanto et al.]{
        R. Priyatikanto,$^{1}$\thanks{E-mail: rhorom.priyatikanto@brin.go.id (RP)}
          E. S. Mumpuni,$^{1}$
          T. Hidayat,$^{2}$
          M. B. Saputra,$^{3}$
          M. D. Murti,$^{3}$
          A. Rachman$^{4}$ and
          C. Y. Yatini$^{1}$
   \\
    $^{1}$Research Center for Space, National Research and Innovation Agency, Bandung 40173, Indonesia\\
    $^{2}$Bosscha Observatory and Astronomy Research Division, Faculty of Mathematics and Natural Sciences, Institut Teknologi Bandung, Bandung 40132, Indonesia\\
    $^{3}$Research Center for Computing, National Research and Innovation Agency, Bogor 16911, Indonesia\\
    $^{4}$Timau National Observatory, National Research and Innovation Agency, Kupang 85361, Indonesia
    }

\date{Accepted XXX. Received YYY; in original form ZZZ}

\pubyear{2022}

\label{firstpage}
\pagerange{\pageref{firstpage}--\pageref{lastpage}}
\maketitle

\newcommand{\himawari}{\emph{Himawari-8}}

\begin{abstract}
A new astronomical observatory in southeastern Indonesia is currently under construction. This Timau National Observatory will host a 3.8-metre telescope for optical and near-infrared observations. To support the operation and planning, the characterisation of the site needs to be appropriately performed. However, limited resources and access to the site hindered the deployment of instruments for comprehensive site testing. Fortunately, \emph{in-situ} sky brightness data from the Sky Quality Meter (SQM) have been available for almost two years. Based on the data acquired in 470 nights, we obtain a background sky brightness of $\mu_0=21.86\pm0.38$ magnitude/arcsec$^2$. Additionally, we evaluate the moonlit sky brightness to estimate the atmospheric extinction coefficient ($k$) and level of scattering on site. We find an alleviated value of $k=0.48\pm0.04$, associated with a high atmospheric aerosol content. It is considered regular for an equatorial area situated at a low altitude (${\sim}1300$ masl). By analysing the fluctuation of the sky brightness and infrared images from \emph{Himawari-8} satellite, we estimate the available observing time (AOT) of at least $5.3$ hours/night and the yearly average percentage of usable nights of $66\%$. The monthly average AOT from SQM and satellite data analysis correlate with $R=0.82$. In terms of the monthly percentage of usable nights, the correlation coefficient is $R=0.78$. During the wet season (November-April), the results from SQM and satellite data analysis deviate more significantly, mainly due to the limited capability of Himawari-8 in detecting fragmented low-altitude clouds. According to these results, we expect Timau to complement other observatories greatly.
\end{abstract}

\begin{keywords}
site testing -- atmospheric effects -- methods: data analysis
\end{keywords}

\section{Introduction}
Currently, Indonesia is developing a new observatory in the East Nusa Tenggara Province to modernise the existing astronomical facilities and develop new ones \citep{mumpuni2018}. The site is situated at the Timau mountain ($123.9471^{\circ}$ East, $9.5970^{\circ}$ South, 1300 masl), approximately 75 km northeast of Kupang, the central city on the island. This location was selected as the new observatory site according to its superior astro-climatic characteristics relative to many other places in Indonesia. \citet{hidayat2012} evaluated the clear sky fractions above Indonesia based on 15-year remote sensing data acquired from 1996 to 2010. Compared to other potential places studied, Timau is the best choice. It has ${\sim}70\%$ clear sky fractions a year, and it is located away from a major inhabited area where light pollution affects astronomical observations. Nevertheless, the site is accessible and close to the existing road networks.

A 3.8-metre optical telescope will be the main workhorse for the new observatory. This telescope will be the sister of the Seimei telescope, which is currently in operation at the Okayama Astrophysical Observatory by Kyoto University \citep{kurita2020}. This telescope will have petal-shaped mirror segments arranged on a relatively lightweight truss structure such that the agility of the telescope will be maximised \citep{kurita2009, kurita2010}. A CCD camera capable of simultaneously capturing the sky in PANSTARRS $g$, $r$, and $i$ bands \citep{maruo2021} and a sensitive near-infrared camera with $Y$, $J$, and $H$ filters are expected to be the first generation instruments for the telescope. The telescope is planned to serve as a general-purpose observing facility for the national astronomical community and international collaborators. More specific scientific missions will be formulated soon.

Meanwhile, site characterisation and continuous monitoring are required to support the observatory's operation, planning and further development. However, some technical issues, such as limited resources and access to the site, hindered the deployment and continuous operation of instruments for \emph{in-situ} monitoring. Earlier reports by \citep{akbar2019} mentioned that the sky brightness at Timau is between 19.63 and 22.18 magnitude per arcsecond square, which is sufficiently good \citep{aksaker2020, deng2021}. These values are based on short-term monitoring using Sky Quality Meter \citep[SQM,][]{cinzano2005, fruck2015, sanchez2017}. The acquired data represent sky brightness in foggy and clear sky conditions, so further examinations are necessary. Following that short campaign, we have performed additional  \emph{in-situ} measurements and observations since 2019 to obtain important site characteristics, including night sky brightness, seeing, and some basic weather parameters \citep{saputra2022}.

Among others, the sky brightness becomes the most interesting one because it can be used to derive several essential parameters related to the site characteristics. \cite{cavazzani2020} proved that the fluctuation of sky brightness data during the night could infer cloud cover. Furthermore, the sky condition can either be categorised as photometric, spectroscopic, or overcast. In line with that, \citet{priyatikanto2020a} trained a random forest model to perform a similar classification task based on some statistical figures extracted from the SQM data. Additionally, the variability of moonlit sky brightness can also be used to estimate the scattering and extinction parameters to a certain degree of accuracy \citep{yao2013, priyatikanto2020b}. These methods can be applied to our new dataset to further characterise the Timau National Observatory.

This paper presents the recent measurement of the sky brightness at the Timau site for more than one year. Besides the general statistics and variability patterns of the night sky brightness, we also estimate the available observing time (AOT), percentage of usable nights, extinction coefficient and scattering parameters based on the SQM data. Even though the estimates are indirect and established from limited \emph{in-situ} data, those parameters complement our existing knowledge about the site.

\section{Data}
\subsection{\emph{In-situ} sky brightness data}
Since May 2020, we have placed and operated a Sky Quality Meter (SQM) to measure sky brightness level over the Timau National Observatory. This instrument consists of a sensitive photodiode equipped with an optical filter to effectively gather photons with wavelengths of 350--600 nanometers \citep{cinzano2005, fruck2015}. Utilising a converging lens in the SQM LU-DL narrows the field-of-view of this instrument down to 20 degrees (full-width at half-maximum) such that the contamination from the surrounding regions in the sky is minimised. Due to the limitation of power supply at the site, the SQM was operated using DC batteries while the data was logged internally before periodical acquisition. We recorded the data that includes the time of measurement both in local and universal time, the ambient temperature in degrees Celsius, and the sky brightness expressed in magnitude per square arcsecond (mpsas).

The SQM was directed to the zenith and was kept inside a container with an acrylic window in the front to measure the sky brightness continuously. This window introduced a 0.24 mpsas offset, which was corrected before further processing and analysis. A degrading window introducing more offset was not identified in the data acquired. Construction work was performed since 2019, so the instrument needed to be located away from the construction site. The work was mainly carried out during the day so that glare was not an issue.

The operation of the SQM commenced in May 2020. Here, we analysed the data acquired by the end of 2021. We also captured a couple of all-sky images from the site using a DSLR camera with a fish-eye lens. In total, there are ${\sim}360,000$ rows of one-minute-cadence data associated with 470 observational nights. During the rainy season, especially from February to April, we could not visit the site for battery replacement. Consequently, there are no data for that period. Tab. \ref{tab:data} summarises some relevant characteristics of the SQM data we used.

The data acquired from Timau are then pre-processed to complement them with the lunar phase data and to identify invalid measurements. The erroneous and invalid data were identified using the Random Forest classifier described by \citet{priyatikanto2020a}. Basically, the classifier characterised the nightly data according to the general statistics and several measures of fluctuation. The classifier assigned the nightly data into six possible classes, namely: (1) peculiar, (2) overcast, (3) cloudy moonlit, (4) clear moonlit, (5) cloudy moonless, and (6) clear moonless. Previously trained using data from several sites in Indonesia, the classifier identified outliers or peculiar data as the ones with unreasonable deviation from the global mean. This class of data shall not be included for further statistical analysis. Moonlit data were identified based on the lunar phase at midnight, while the dichotomy of cloudy and clear sky was mainly based on the inter-quartile range, median absolute deviation, and the percentage of data that deviates more than two mpsas from the median value.

\begin{table}
    \caption{Summary of the SQM and satellite data used in this study.}
    \label{tab:data}
    \centering
    \begin{tabular}{lccc}
    \toprule
    & SQM & \multicolumn{2}{c}{\himawari} \\
    Characteristics & & Band 8/IR3 & Band 13/IR1 \\
    \midrule
    Wavelength ($\mu$m) & $0.35-0.60$ & $5.9-6.6$& $10.1-10.7$ \\
    Spatial resolution & -- & \multicolumn{2}{c}{$0.05^{\circ}$ (${\sim}5.6$ km)} \\
    Time resolution & 1 min & \multicolumn{2}{c}{1 hour} \\
    Time range & 2020-05-20 & \multicolumn{2}{c}{2020-01-01}\\
    & to 2021-12-22 & \multicolumn{2}{c}{to 2021-12-31}\\
    \bottomrule
    \end{tabular}
\end{table}

\subsection{Satellite data}
As a complement, we used satellite images acquired by a meteorological satellite {\himawari} \citep{bessho2016}. This satellite observes the Earth from a geostationary position above $140.7^{\circ}$ E longitude using Advanced Himawari Imager (AHI). It captures full-disk images of the Earth at 16 channels covering optical, near-infrared and infrared windows. Compared to the imager of the \emph{Multi-functional Transport Satellite} (\emph{MTSAT}) as its predecessor, AHI has a higher sensitivity and spectral resolution at some channels. Among others, images from channels 8 (6.2 microns) and 13 (10.4 microns) are essential for detecting clouds. The former channel coincides with the water absorption band such that the brightness temperature at this channel indicates the humidity in the middle-to-upper troposphere. The latter channel can measure the amount of water vapour at lower altitudes. Tab. \ref{tab:data} summarises some relevant characteristics of the {\himawari} data we used.

\begin{figure}
    \centering
    \includegraphics[width=\columnwidth]{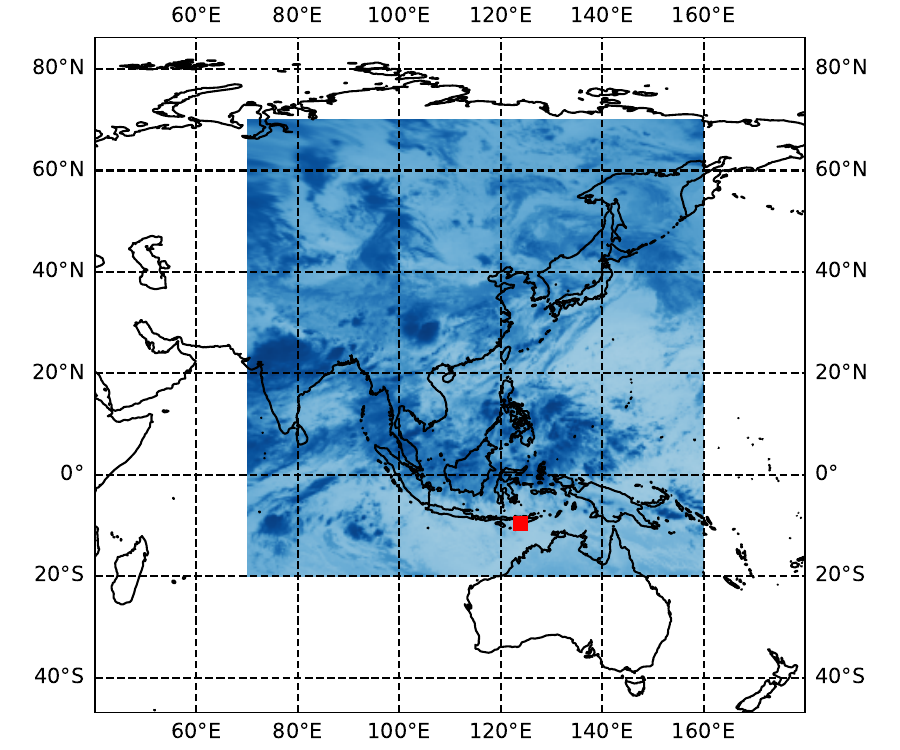}
    \caption{The regional map with a superimposed infrared IR1 image from {\himawari} acquired on 2020-12-20 at 13:00 UT. The location of Timau National Observatory is marked by a red square.}
    \label{fig:map}
\end{figure}

\begin{figure*}
    \centering
    \includegraphics[width=\textwidth, trim=0 0 0 40, clip]{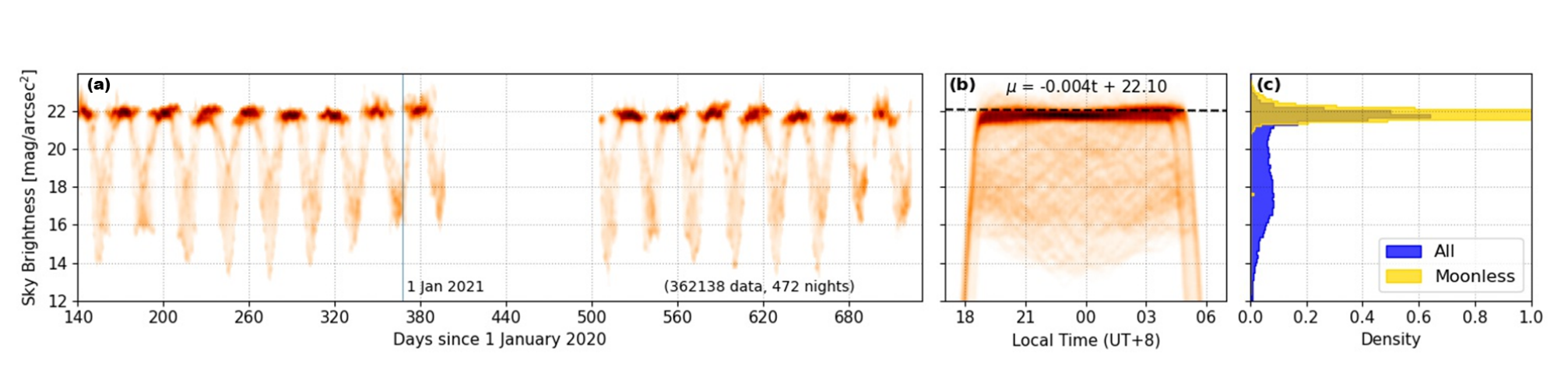}
    \caption{Density plot of the sky brightness over Timau as a function of the number of days since 1 January 2020 (a) and as a function of the local time (b). One-dimensional distribution of the sky brightness is also presented (c).}
    \label{fig:sqm_dist}
\end{figure*}

We downloaded hourly images of the {\himawari} bands 8 and 13, which were acquired from 1 January 2020 to 31 December 2021. Kochi University\footnote{\url{http://weather.is.kochi-u.ac.jp/archive-e.html}} provides archive of images from Himawari 8 and its predecessors, among which geo-coordinated data covering the area from $70^{\circ}$ E to $160^{\circ}$ E and from $20^{\circ}$ S to $70^{\circ}$ N (e.g., see Fig. \ref{fig:map}). The images are stored as $1800\times1800$ pixels$^2$ of 8-bit portable grey maps (PGMs) with a spatial resolution of $0.05^{\circ}$. In the archive, band 8 images are marked with the IR3 suffix as they resemble the IR3 images from the MTSAT. In the same way, band 13 images are marked with the IR1 suffix. Henceforth, we will use IR3 and IR1 to represent band 8 and band 13 images. Fig. \ref{fig:map} displays an example of a Himawari 8 image covering a vast area of interest. The cloud cover over the Timau site was estimated based on the values of a sample of $2\times2$ pixels$^2$ centred on the defined geographic coordinate. This sample is smaller than the $5\times5$ pixels$^2$ used in \citet{hidayat2012} because we are currently interested in identifying clouds around the zenith point. Therefore, we can perform an apple-to-apple comparison with the output from the SQM data analysis. Moreover, we are interested in the clear sky in the time domain (duration), while the previous study regarded the clear sky fractions in a spatial sense. Besides those images, Kochi University also provides a table to convert $0-255$ pixel values to the brightness temperature at a certain band. In general, the higher pixel value corresponds to the lower brightness temperature though the relationship is not linear.

\section{Sky brightness over Timau}
\subsection{General statistics}
Based on the data acquired in 470 nights, we established multi-dimensional distribution of sky brightness over the site. This acted as the basis for further exploratory analyses. Fig. \ref{fig:sqm_dist} displays the density plots of the data in two different temporal spaces (day versus local time) and the aggregation over the time domain.

\begin{figure}
    \centering
    \includegraphics[width=\columnwidth]{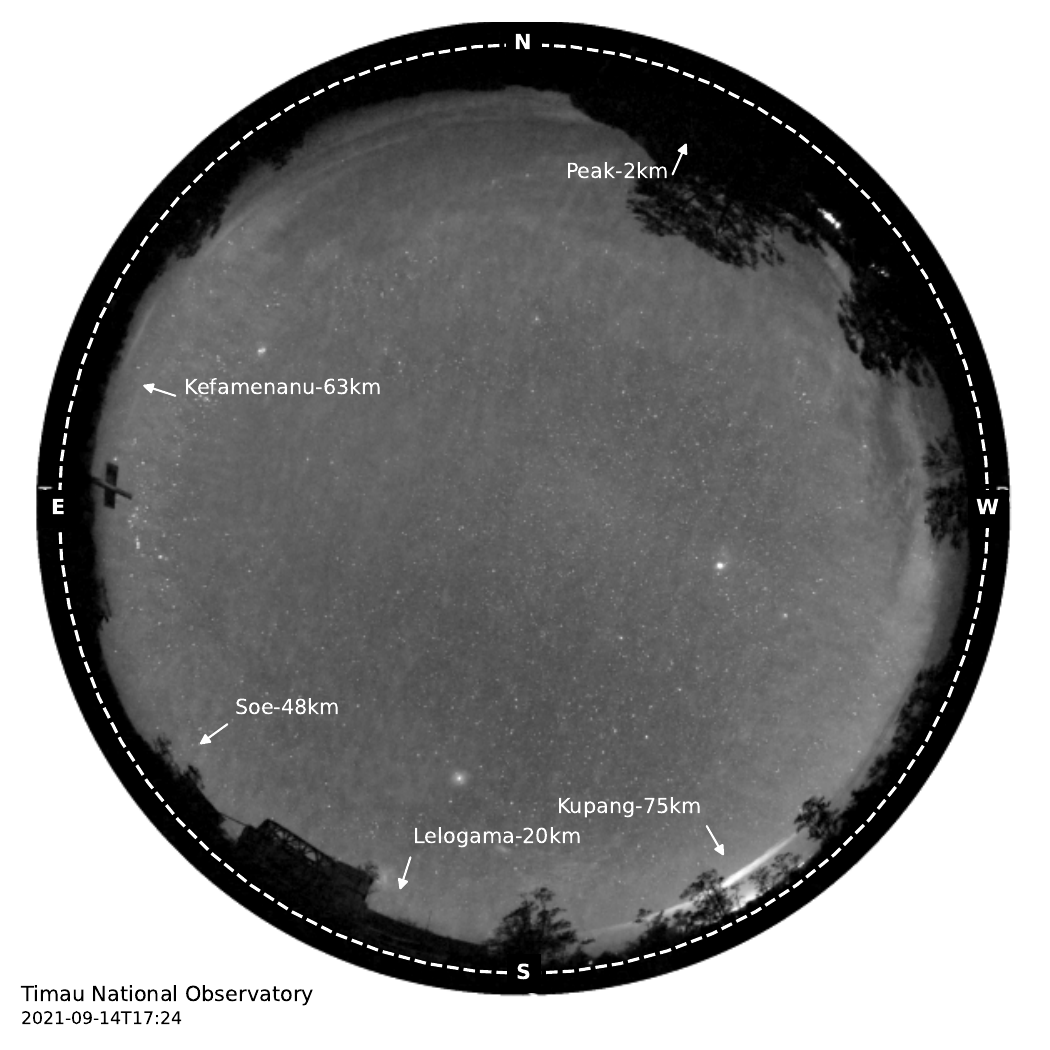}
    \caption{All-sky image acquired at Timau on 2021-09-14 shows a starry dark sky with little light pollution at the south-south-west horizon, which is coming from Kupang City.}

    \label{fig:asi}
\end{figure}
In the first density plot (Fig. \ref{fig:sqm_dist} (a)), we can see densely packed data points at around $22.0$ mpsas that correspond to the night sky brightness over Timau and the periodic variability of the sky brightness. Modulation with the lunar phase progression is seen in the plot, and the local minima vary from 13 to 16 mpsas. Those minima are associated with the brightest sky during the full moon. Periodic changes in the Sun-Moon distance, the Earth-Moon distance, and the variation in the topocentric position of the Moon are the significant factors that influence the observed variation of the moonlit sky brightness, including the minima. The Sun-Moon distance governs the intrinsic brightness of the Moon, while the observed brightness is a function of the Moon’s distance from the observer and the phase \citep{krisciunas1991}. Lastly, the angular distance between the Moon and the field-of-view of the SQM determines the amount of scattered (and direct) moonlight recorded by the instrument.

The second exciting feature of the plot is the positive deviation of night sky brightness ($\mu>22.0$ mpsas), which is more common during the wet season around January. As mentioned in the literature \citep{jechow2016, posch2018, jechow2019}, cloud cover alters the sky brightness over the urban and rural areas differently. In artificially lit urban areas, clouds brighten the sky up to 15 times. On the other side, the pristine sky over remote regions can be darkened by passing clouds that attenuate the light from extraterrestrial sources. The one-magnitude positive deviation observed in Fig. \ref{fig:sqm_dist} agrees with previously reported behaviour \citep{posch2018, jechow2019}.

Next, the middle panel of Fig. \ref{fig:sqm_dist} shows the night sky brightness as a function of local time. This plot is also known as the jellyfish diagram \citep{posch2018}. However, the small variations of sunset and sunrise time at Timau produce ‘tentacles’ that are more concentrated on both ends of the plot than the diagram created using data from regions at higher latitudes. Those parts represent the twilight signals that depend on the altitude of the Sun. It is noteworthy that the maximum difference between the earliest and the latest sunset/sunrise time at Timau is less than 50 minutes. The middle part of the plot tells us about the night sky brightness both during the moonless (middle-top) and the moonlit night (see, for example, \citet[]{bara2019}). There is no significant darkening overnight, as indicated by the flat top of the jellyfish diagram. To ensure this notion, we extracted the 90\textsuperscript{th}-percentile of the hourly night sky brightness from 20.00 to 04.00 local time. A linear function was then fitted to those figures to get a simple model of systematic darkening during the night. However, the observed declining rate of the night sky brightness over Timau is just $0.004$ mpsas/hour which is negligible. This is reasonable since systematic darkening rarely occurs in rural areas, while in urban areas, the darkening can be $0.14$ mpsas/hour \citep{posch2018}. In line with that, \citet{herdiwijaya2020} found a ${\sim}0.7$ mpsas  deviation between pre- and post-midnight sky brightness over Bosscha Observatory, where light pollution from surrounding urban areas is inevitable.

Abrupt changes in the sky brightness overnight are more likely to be linked to human activities, e.g., reduction of outdoor lighting after midnight or the settlement of aerosol at low altitudes. On the other hand, systematic changes in the natural sky brightness are associated with the variation of airglow emissions. The emission from excited OH molecules is believed as the dominant source of airglow emissions in optical and infrared windows. Based on the spectroscopic observations using the Sloan Digital Sky Survey, \citet{hart2018} presented a decrease in OH airglow emission after sunset until midnight, followed by a moderate increase before sunrise. The ratio between maximum airglow emission intensity at dusk/dawn to the minimum value at midnight depends on the emission mode, but the typical ratio is less than $1.2$. Assuming that the airglow emission contributes to ${\sim}50\%$ of the natural night sky brightness \citep{masana2021}, we can expect the difference of $0.1$ mpsas overnight. However, this natural darkening is not observed in the density plot presented in Fig. \ref{fig:sqm_dist}.

The aggregate one-dimensional distribution of the sky brightness over Timau can be seen in Fig. \ref{fig:sqm_dist}(c). To extract the statistical figures representing the night sky brightness at the site, we selected a subset of data acquired when the Moon’s altitude was below $-5^{\circ}$ and the Sun’s altitude was below $-18^{\circ}$. Univariate distribution established from this subset is more concentrated around the average sky brightness of ${\mu_0}=21.86$ mpsas with a standard deviation of $\sigma_{\mu}=0.38$. These statistical figures emphasise that Timau has a pristine sky which is suitable for astronomy. It is undoubtedly reasonable since this site is a remote place, away from municipalities. Kupang, the major city in West Timor, is located 75 km away from the site and only contributes to the minor light pollution at the south-south-western horizon. This was identified from the all-sky image acquired at the Timau site (Fig. \ref{fig:asi}). Kefamenanu and Soe are two smaller towns situated 60 and 50 km from the site, respectively. As a supplement, Kupang City has a maximum population density of 10,000 km$^{-2}$ while Kefamenanu and Soe, respectively have 1000 and 2000 residents per square kilometre \citep{worldpop}. Within a 20 km radius of the site, there are several villages with minimum access to electricity, and hence light pollution is suppressed to the minimum. Some residents in those villages use solar panels to generate electricity for indoor lighting.

\subsection{Moonlit sky brightness}
\begin{figure*}
    \centering
    \includegraphics[scale=0.8]{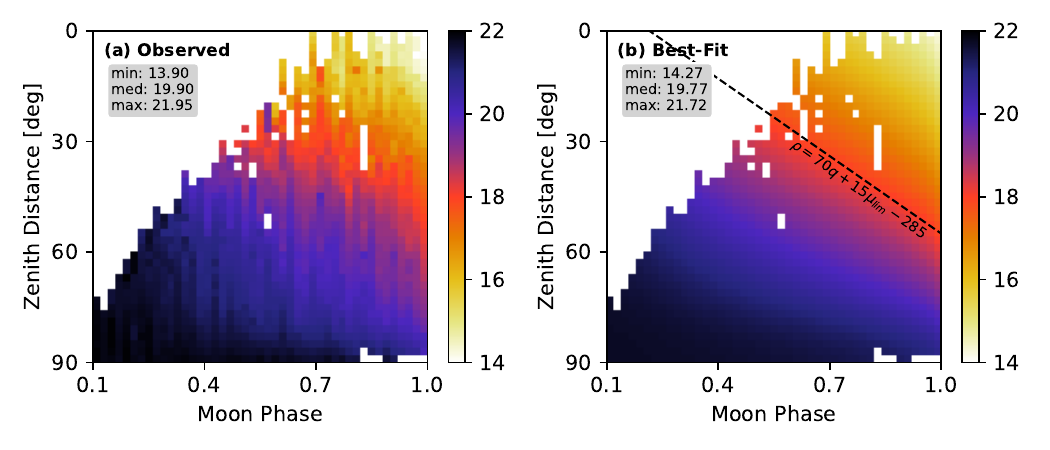}
    \caption{Observed (a) and calculated (b) sky brightness over Timau as a function of phase and zenith distance of the moon.}
    \label{fig:ks91}
\end{figure*}

The monthly cycle of sky brightness can roughly describe the atmosphere above the site. \citet{krisciunas1991} presented a semi-empirical model of moonlit sky brightness that depends on the local extinction coefficient and the parameters related to the Rayleigh and Mie scattering. Moonlit sky brightness measured using SQM had been exploited to find the suitable parameters for the model \citep{yao2013, priyatikanto2020b}. In brief, the model is described with the following equations:
\begin{align}
    B_0 &= 34.08e^{20.723-0.921\mu_0}, \\
    I &= 10^{-0.4(3.84+0.026|\alpha|+4\times10^{-9}\alpha^4)}, \\
    f_R &= 10^{5.36}(1.06+\cos\rho^2), \\
    f_M &= 10^{6.15-0.025\rho}, \\
    f &= P_Rf_R + P_Mf_M, \\
    B &= fI\left(10^{-0.4kX_{\text{moon}}}\right)\left(1-10^{-0.4kX}\right), \\
    \mu &= \mu_0-2.5\log\left(\dfrac{B}{B_0}+1\right),
\end{align}
where $B_0$ and $\mu_0$ represent moonless sky brightness in linear (nanoLamberts) and logarithmic scale (mpsas), respectively, $I$ denotes Moon brightness at specified phase angle $\alpha=\cos^{-1}(2q-1)$, with $q$ represents the Moon phase or the fraction of illumination. Next, $k$ represents the extinction coefficient, $f_R$ and $f_M$ are the Rayleigh and Mie scattering functions, while $P_R$ and $P_M$ are the scaling factor of those functions. The airmass in the SQM direction $X$ is a function of zenith distance $Z$, which is $0^{\circ}$ for our case, while the airmass in the Moon direction ($X_{\text{moon}}$) is variable. The observed moonlit sky brightness ($B$ or $\mu$) depends on the angular distance $\rho$ between the Moon and the SQM.

To obtain the most suitable parameters for the model, we mapped the observed sky brightness onto a two-dimensional space where the Moon’s zenith distance and phase become the axes. For practical purposes, the data were binned into a grid of size $0.02$ in the Moon phase and $2^{\circ}$ in zenith distance. Subsequently, we computed the average sky brightness in each bin as the basis of the model fitting. As in \citet{priyatikanto2020b}, we utilized Affine Invariant Markov Chain Monte Carlo (MCMC) Ensemble implemented in the \textsc{emcee Python} package \citep{foreman2013} to randomly sample the possible parameters ($\mu_0\in(15,23), k\in(0,2), P_R\in(0,15), P_M\in(0,15)$) such that the optimum parameters can be obtained. The sampler started with 100 sets of parameters distributed over the parameter space and rolled over 5000 iterations to reach the best solution with the maximum likelihood. The parameter uncertainties were also estimated. In the end, we obtained $\mu_0=21.72\pm0.03$, $k=0.48\pm0.04$, $P_R=0.04^{+0.05}_{-0.03}$, $P_M=9.82^{+0.26}_{-0.29}$ as the best-fit parameters. To check for possible differences in the atmospheric characteristics during dry and wet seasons, we separately fitted the data acquired during May-September (dry) and October-December (wet). However, we found no evidence for such differences. Fig. \ref{fig:ks91} shows the observed sky brightness as a function of the Moon phase and zenith distance, both from observation and modelling. The root-mean-square error for that model is $0.37$ mpsas.

Exploiting the moonlit sky brightness model is not the best way to get the observational characteristics of the site. However, this approach provides the first order of estimate while more precise photometric data is not available yet. Moreover, the optimum parameters arguably make sense. Firstly, the background sky brightness $\mu_0$ agrees with the global statistics presented. Secondly, we obtained the extinction coefficient of $0.49$, which is relatively high but comparable to the value obtained at the Bosscha Observatory, Indonesia ($6.8^{\circ}$ S). In principle, the extinction coefficient depends on the observational wavelength. For instance, the V-band extinction coefficient at Mauna Kea is $0.11$ \citep{button2013}, at Cerro Paranal is $0.12$ \citep{patat2011}, at Dome C Antarctica is $0.13$ \citep{chadid2019}, and at Gaomeigu China is $0.20$ \citep{hu2011}. Based on the moonlit sky brightness modelling, \citet{yao2013} derived $k=0.23$ for Xinglong Station China. At Bosscha Observatory, \citet{malasan2020} reported the extinction coefficient between $0.30$ and $0.88$ based on the spectroscopic observation ($4200-7000$ \AA) in 2019. That quoted value is more than double the coefficient derived for the same site in 1993. At $5500$ \AA, the recent extinction coefficient is ${\sim}0.45$, comparable to the one obtained for the Timau site study. Disregarding the distinction between the Bosscha Observatory in the western part of Indonesia and the Timau in the east, these two sites are still in the tropical archipelago.

Next, the total atmospheric extinction is known to be the sum of extinction by small particles through the Rayleigh scattering process, by the aerosols via the Mie scattering process, and by the absorption of ozone particles \citep{hayes1975}. The elevated extinction coefficient in Indonesia is likely associated with the high aerosol content in the region.  The low altitude of the sites ($\sim$1300 masl for both locations) also makes the aerosol component more prominent.

A significantly high contribution of the Mie scattering process in the atmosphere above Timau is also reflected by the extreme value of $P_M=9.82$. In contrast, the scaling factor for the Rayleigh scattering is just $P_R=0.04$. These values mean that the amount of scattering by the aerosols above Timau is ten times higher than that above Mauna Kea, where the semi-empirical scattering function was calibrated in \citet{krisciunas1991}. On the other hand, the contribution from the Rayleigh scattering is suppressed.

\subsection{Aerosols above Indonesia}
Aerosol content in the atmosphere plays an important role in astronomical observation as it affects transparency \citep{lombardi2008, aksaker2020} and determines the amount of sky glow scattered in the atmosphere \citep{sanchez2020, kocifaj2020}. According to the MODIS Aerosol Product \citep{sayer2014, aksaker2020}, the atmosphere above Indonesia has a moderate aerosol optical depth (AOD) such that the atmospheric transparency is certainly degraded. In line with this, in-situ measurements under the AERONET framework \citep{giles2019} obtained a typical value of AOD at 500 nm of 0.13 from 6 stations outside Java Island. Urban and industrial activity or biomass burning may elevate the aerosol content significantly \citep{kusumaningtyas2022}. Seasonal variation of AOD in Indonesia is prominent as the rise of humidity and rainfall during the wet season reduces the aerosol content. Meanwhile, the dry season is often accompanied by biomass burning that boosts the aerosol content. The western part of Indonesia (e.g., Kalimantan and Sumatra islands) occasionally becomes the aerosol source during the dry season. Smoulder burning in the peat land produces smoke that contains more organic carbon than black carbon. It is known that organic carbon is less absorbent than black carbon \citep{eck2019}. The smoke is transported northeast by the monsoonal wind \citep{xian2013}. So, it is fortunate that the burning season in the west will not directly affect the atmospheric condition in the Timau site.

The AOD characteristics over Timau can be approximated using the AERONET data obtained at Kupang station\footnote{\url{https://aeronet.gsfc.nasa.gov/new_web/aerosols.html}}. In 2020 and 2021, the average AOD over this station was around $0.13$. In general, the AOD is reduced during the wet season, as in the other stations. However, the aerosol composition over Kupang is somewhat different compared to the characteristics obtained in the remaining stations. In Kupang, the average Angstr{\"o}m exponent (AE) of ${\sim}1.1$ means that the fine particle is still the dominant component of the aerosol \citep{schuster2006}, but larger particles like dust and sea salt can be found. As a comparison, the AEs from other stations are ${>}1.2$ \citep{kusumaningtyas2022}. Because the monsoonal wind is northwest directed during the dry season, the dusty aerosol from Australia likely affects the aerosol composition in Kupang.

These characteristics can be related to the results from the SQM data. Firstly, AERONET measured the AOD over Kupang as $0.13$ while the AOD over Mauna Loa (Hawaii) is about $0.02$. Assuming that AOD over Timau is not significantly different to Kupang and that AOD over Mauna Kea is indifferent to Mauna Loa, we can accept that the aerosol scattering over Timau is approximately ten times more intense.
Secondly, we obtained an extinction coefficient of $k=0.48$. At an airmass of 1, this value translates to an atmospheric optical depth of $\tau=0.4\ln(10)k=0.44$, which is more than three times the AOD over Kupang. Then, we expect more intense scattering by atmospheric molecules to explain the extinction coefficient derived. This expectation contradicts the statement that the contribution of Rayleigh scattering is suppressed. More direct observations are required to unravel this issue.

\section{Usable nights at Timau}
\subsection{Clear sky identification using SQM data}
\newcommand{\sigmaobs}{\sigma}

Cloud cover alters the sky brightness that is measured by the SQM. The change can either be gradual or erratic, depending on the structure of the cloud. The change can also have different directions depending on the site's location relative to the source of artificial light pollution. In the urban region where light pollution is imminent, cloud cover makes the sky brighter as it scatters more artificial light back to the ground. On the other hand, the pristine sky tends to be darker when the cloud blocks the view to the starry night sky. The magnitude of the changes also modulates with the lunar phase. These principles are used to identify cloud covers based on the temporal fluctuation of sky brightness \citep{cavazzani2020}. In this approach, the standard deviation ($\sigmaobs$) is computed in a time bin and compared to a certain threshold ($\sigma_{\text{thres}}$) that varies with the moon phase $q$ and depends on the light pollution level of the site. Clear sky conditions are identified when $\sigmaobs < \sigma_{\text{thres}}$. \citet{cavazzani2020} described a linear relationship between $q$ and $\sigma_{\text{thres}}$. They referred to the global map of light pollution \citep{falchi2016}\footnote{\url{https://lightpollutionmap.info}} to estimate $\sigma_{\text{thres}}$. However, this prescribed $\sigma_{\text{thres}}$ leads to an overestimation of the clear sky in our dataset, especially during the early phase of the Moon. Thus, we used the following quadratic function to define the threshold:
\begin{equation}
\sigma_{\text{thres}} = 0.012 + 0.112q^2.
\label{thr}
\end{equation}
This empirical function was obtained by inspecting the $x$-th percentile of the $\sigma$ as a function of the Moon phase (binned). The value of $x$ influences the annual percentage of usable nights, but it does not affect the seasonal variation. We used $x=70$ to get a result that is comparable to the satellite-based result. Using Eq. \ref{thr}, the percentage of data associated with the clear sky conditions is independent of $q$. In other words, this quadratic function dictates a larger $\sigma$ for a cloudy moonlit sky.

In practice, we computed $\sigmaobs$ using a 12-minute bin and computed a single $\sigma_{\text{thres}}$ for each bin. We performed the computation on the data acquired from 19.04 to 04.40 local time. This defined window covers the 9.6-hour nighttime, from the astronomical dusk to the astronomical dawn.

After identifying clear skies, we computed the daily available observing time (AOT), which is the total clear hours from dusk to dawn. We categorised the nights into photometric, spectroscopic, or overcast nights. A photometric night has an AOT of more than 6 hours, while a spectroscopic night has clear hours between 2 and 6. Otherwise, the night is flagged as overcast. Finally, we calculated the percentage of usable nights at a certain period. It is defined as the total number of photometric and spectroscopic nights divided by the number of nights (with available data).

\subsection{Clear sky identification using satellite data}
Following the work of \citet{hidayat2012}, we employed a simple thresholding method to identify clouds over the site. The basis of this method is that the brightness temperature measured at the IR3 channel correlates with the middle-to-upper tropospheric humidity \citep[$UTH$,][]{erasmus2002}. According to \citet{da2015}, this channel works effectively to capture the atmospheric temperature at the altitude of 6--12 km. On the other hand, the brightness temperature at IR1 can be used to identify clouds at lower altitudes as it captures the conditions near the surface. At night, the effective altitudes are slightly elevated due to the change in the temperature profile \citep{cavazzani2015}. A cloud can be formed when the humidity exceeds a certain threshold or when the brightness temperature drops below a specific level.

\begin{figure}
    \centering
    \includegraphics[width=\columnwidth]{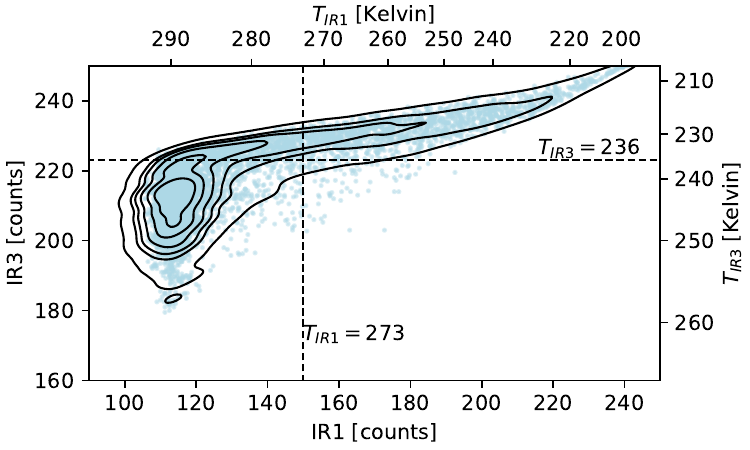}
    \caption{Density plot of the pixel values in IR1 and IR3 bands. The associated brightness temperatures are also shown as the additional axes. The dashed lines mark the thresholds for clear sky identification.}
    \label{fig:mtsat}
\end{figure}

Fig. \ref{fig:mtsat} shows the density plot of the IR1 versus IR3 pixel values and the relevant brightness temperatures. We can see a turn-off in the plot related to the threshold we can use to identify cloud cover in the satellite image. Based on this figure, we defined $T_{\text{IR1}}=273$ K and $T_{\text{IR3}}=236$ K as the most suitable thresholds for cloud identification. The sky at a certain hour was categorised as clear if the brightness temperatures $T_{\text{IR1}}>273$ K and $T_{\text{IR3}}>236$ K. The former threshold is similar to the one used in \citet{hidayat2012} while the threshold for $T_{\text{IR1}}$ is slightly lower. Following the empirical relation from \citet{erasmus2002}, the associated $UTH$ at this brightness temperature is 100\%. The adopted temperature thresholds are reasonable considering the fact that the dew point temperature decreases by altitude. We also checked the vertical profile of the dew point temperature at midnight from the ERA5 dataset \citep{hersbach2020}. At the altitude of 0--3 km, the dew point is 270--290 K. A lower dew point temperature of 210--250 K is expected at the altitude of 6--12 km. If we refer to Fig. 5, the identification of clear sky condition and the overall percentage of usable nights are not sensitive to the change of IR1 threshold as the shift of this threshold does not significantly change the proportion of the data inside the lower left quadrant. However, the identification is more sensitive to the choice of threshold in IR3. Thus, we selected a threshold that minimised the mean squared error between the expected monthly percentage of usable nights and the results from both SQM data analysis and \citet{hidayat2012}.

Similar to SQM data analysis, we counted the total of clear sky hours per night and noted the usable night accordingly. The implementation of this method in the {\himawari} data yielded the percentage of usable nights in 2020-2021 of 66\%, which is compatible with the result of \citet{hidayat2012}. Based on the 15-year statistics, they concluded that the average fraction of usable nights at Timau is 63.3\% with a standard deviation of 6.3\%.

\subsection{Temporal variability}
\begin{figure}
    \centering
    \includegraphics[width=\columnwidth]{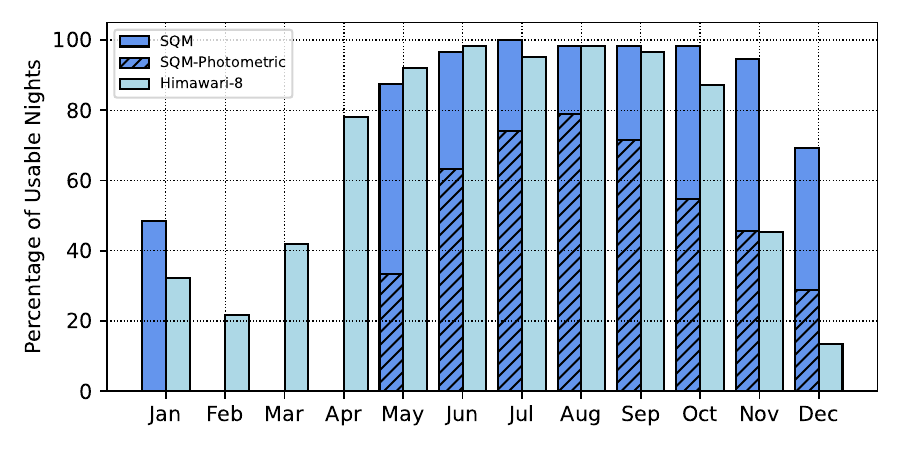}
    \caption{The seasonal variation of the usable night according to the SQM and {\himawari} data. The fraction of photometric nights can be estimated using SQM data.}
    \label{fig:monthly}
\end{figure}

Most of the Indonesian archipelago experience a wet/rainy season that starts from around October, while the commencement of the wet season in the eastern part of Indonesia is a bit later \citep{hamada2002}. The chance of getting usable clear nights is lower during the wet season that lasts from November to April, whereas the fraction of usable night hikes to the maximum in May and declines in October. This seasonal variation is seen in Fig. \ref{fig:monthly} where the percentages of the usable nights were derived from the SQM and {\himawari} data. The fractional values associated with the photometric nights are also presented as hatched bars in Fig. \ref{fig:monthly}.

In general, the monthly average of the AOT from SQM and {\himawari} data correlate to each other with a correlation coefficient of $R=0.82$. In terms of the monthly percentage of usable nights, the correlation coefficient is $R=0.78$. During the dry season (May-October), the percentages of usable nights from SQM agree well with the ones derived from {\himawari} data. The differences between those two solutions are around $4\%$. Based on the SQM data, we estimated the occurrence of photometric nights with total clear hours of more than 6 hours to be around $40\%$. The deviations between the results from SQM and {\himawari} are more prominent during the wet season. Our analysis of the SQM data leads to the percentage of usable nights of ${>}45\%$ while the satellite-based figures are ${<}45\%$. These differences are significantly larger than the results from \citet{cavazzani2020}. For La Silla and Asiago sites, the SQM-based and satellite-based clear sky percentages agree with ${\lesssim}10\%$ differences. Distinct characteristics between the tropical site we evaluate and the subtropical ones analysed by \citet{cavazzani2020} may be the foremost reason why our SQM-based results deviate from the satellite-based ones.

\begin{figure}
    \centering
    \includegraphics[width=\columnwidth]{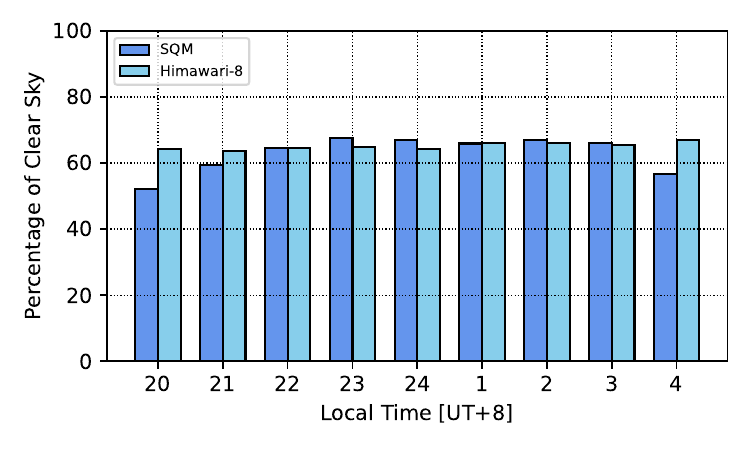}
    \caption{The daily variation of the clear sky percentages from the SQM and satellite data outside Februuary-April.}
    \label{fig:daily}
\end{figure}

\begin{table*}
    \newcommand{\tulis}[6]{#1 & $#2$ & $#3$ & $#4$ & $#5$ & $#6$}
    \newcommand{\tulise}[3]{& $#1$ & $#2$ & $#3$ \\}
    \caption{Monthly average of the AOT and the percentage of usable nights derived from the SQM and satellite data. For the SQM-based, the percentages of spectroscopic and photometric nights are also provided. The number of nights analysed is presented in column $n$. The yearly aggregated values (total $n$, average AOT and usable night) are provided in the bottom row.}
    \label{tab:results}
    \centering
    \begin{tabular}{ccccccccc}
    \toprule
    Month & \multicolumn{5}{c}{SQM-based} & \multicolumn{3}{c}{Satellite-based} \\
    & $n$ & AOT & \% spec & \% phot & \% usable & $n$ & AOT & \% usable \\
    \midrule
    \tulis{Jan}{31}{2.3}{45}{0}{45}\tulise{56}{1.5}{31}
    \tulis{Feb}{-}{-}{-}{-}{-}\tulise{55}{1.5}{22}
    \tulis{Mar}{-}{-}{-}{-}{-}\tulise{62}{2.7}{42}
    \tulis{Apr}{-}{-}{-}{-}{-}\tulise{59}{6.1}{78}
    \tulis{May}{24}{5.0}{62}{21}{83}\tulise{62}{7.7}{92}
    \tulis{Jun}{60}{5.9}{55}{37}{92}\tulise{60}{8.1}{95}
    \tulis{Jul}{62}{6.7}{52}{47}{98}\tulise{62}{8.1}{95}
    \tulis{Aug}{62}{6.8}{40}{56}{97}\tulise{62}{8.6}{98}
    \tulis{Sep}{60}{6.5}{55}{43}{98}\tulise{60}{8.1}{97}
    \tulis{Oct}{62}{5.9}{65}{34}{98}\tulise{62}{7.1}{91}
    \tulis{Nov}{57}{5.6}{53}{33}{86}\tulise{55}{2.9}{64}
    \tulis{Dec}{52}{3.8}{40}{19}{60}\tulise{59}{0.8}{39}
    \midrule
    \tulis{Agg.}{470}{-}{-}{-}{-}\tulise{714}{5.3}{66}
    \bottomrule
    \end{tabular}
\end{table*}

The differences can also be observed in the percentage of the clear sky as a function of local time (Fig. \ref{fig:daily}). To make a fair comparison, we computed the hourly clear sky fractions based on the data acquired outside February to April each year. The SQM-based results show an obvious variation of clear sky fractions overnight. Starting from a moderate value of ${\sim}50\%$, the clear sky fraction rises as the night progresses. After reaching a maximum value of ${\sim}65\%$ at midnight, it declines more quickly after 3 a.m. local time. Twilight is unlikely to cause the observed variation since the altitude of the Sun is below $-18^{\circ}$ between 8 p.m. and 4 a.m. The observed asymmetry cannot be associated with the scattered moonlight because its effect should be homogeneous overnight. There should be no preferred hours during which the moonlight affects the identification of a clear sky from SQM data. Moreover, our method relies on the temporal fluctuation of sky brightness, so the gradual brightening by moonlight should not be a problem. Consequently, the observed variation of clear sky fractions overnight is real and likely to be linked to the weather phenomena at low altitudes. Due to this reason, {\himawari} cannot identify the variability of the hourly clear sky fractions. As discussed by \citet{lagrosas2022}, {\himawari} has a limitation in detecting low altitude (${<}2$ km) clouds, especially the small fragments. On some occasions, the occurrences of low-altitude clouds, mainly in the first half of the night, were confirmed during several site visits.

This clear sky identification discrepancy can be associated with two possible causes. Firstly, the thresholds we used for the satellite data are too low. A higher count threshold means assuming a lower brightness temperature and an upper-tropospheric humidity of more than 100\%, which is unlikely. Secondly, at some occasions the presence of a homogeneous opaque cloud alters the night sky brightness without introducing any sensible fluctuation. A reliable ground data (all-sky image or observing log) can provide a settlement for this discrepancy, but such data is not available for the Timau site at this moment.

\begin{figure}
    \centering
    \includegraphics[width=\columnwidth]{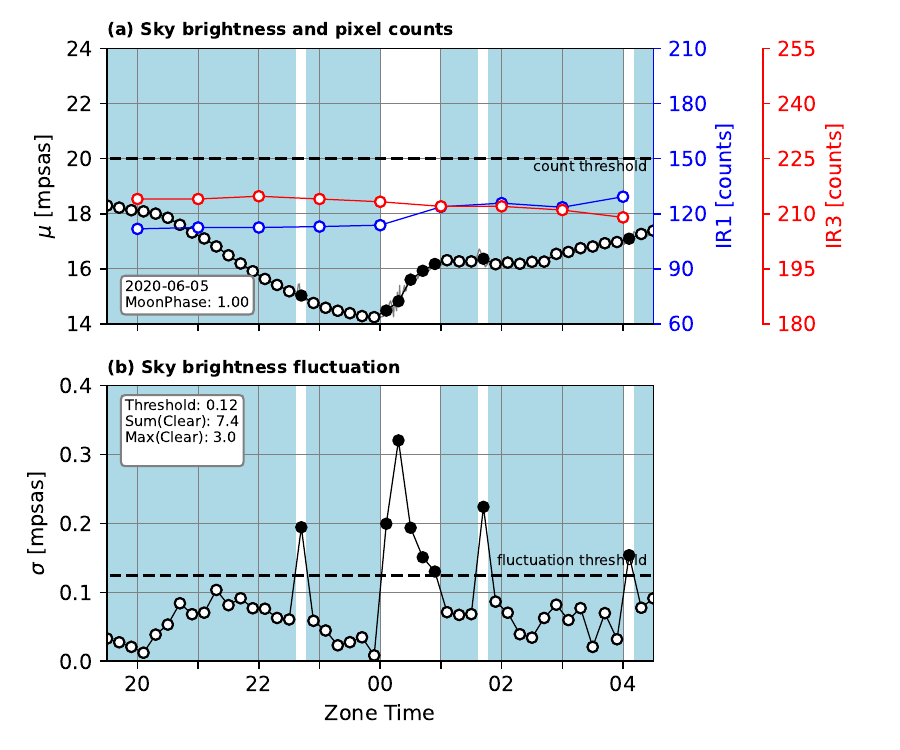}
    \vspace{-2em}
    \caption{Cloud identification on 2020-06-05 based on the fluctuation of night sky brightness and the pixel counts from the IR1 and IR3 {\himawari} images. The fluctuation threshold for the SQM data and count thresholds for the satellite data are indicated by dashed lines. Filled circles indicate cloud covers.}
    \label{fig:cek1}
\end{figure}

\begin{figure}
    \centering
    \includegraphics[width=\columnwidth]{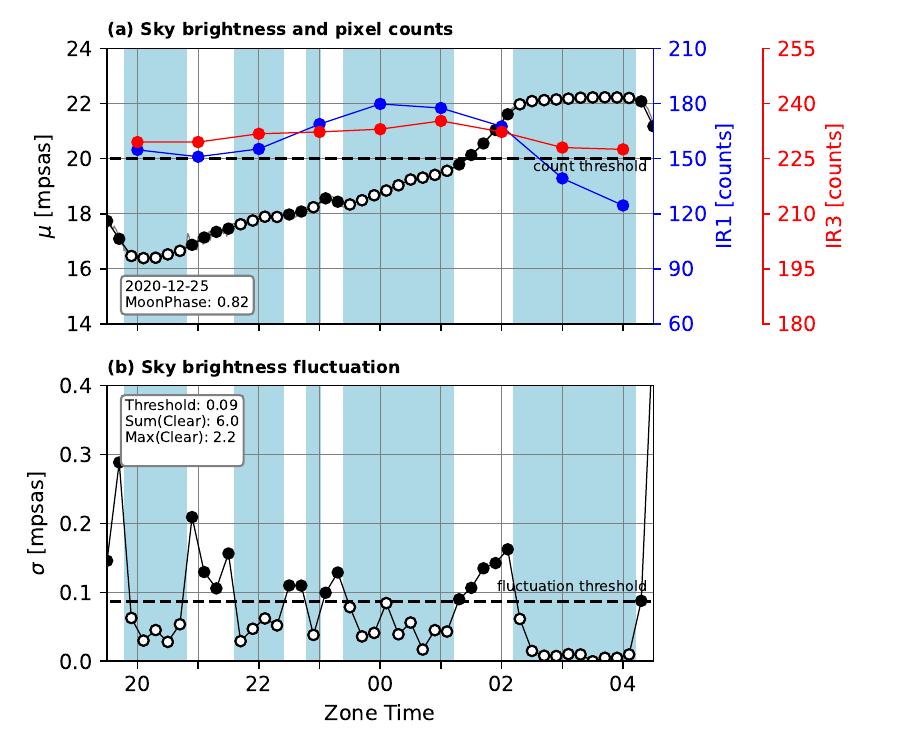}
    \vspace{-2em}
    \caption{Same as Fig. \ref{fig:cek1}, but for the detection on 2020-12-25.}
    \label{fig:cek2}
\end{figure}

To explore more, we performed examinations of the nightly data. Fig. \ref{fig:cek1} and \ref{fig:cek2} indicate the differences between the SQM-based and satellite-based cloud detection on two selected nights (2020-06-05 and 2020-12-25), representing dry and wet seasons. As blue and red dots indicate, the remotely sensed brightness temperatures are relatively stable overnight. In the IR1 channel, the derived brightness temperatures slightly fluctuate overnight with the typical standard deviation of 10 K. In the IR3 channel, the typical variation is even more minor, around 3 K. As a result, the categorisation of the night sky becomes more contrast when we use {\himawari} data. The night sky is identified either as overcast (short AOT) or photometric (long AOT), while the transitional conditions (partially cloudy or spectroscopic night) are rarely identified. This is in line with the finding from \citet{hidayat2012} that the percentage of nights with transitional conditions is only $4.6\%$. In contrast, the SQM data has a higher time resolution such that the variable state of the night sky can be captured in more detail. In Fig. \ref{fig:cek1}, the asymmetric profile of the moonlit sky brightness especially during the lunar sunset is evidence of enhanced aerosol content. In Fig. \ref{fig:cek2}, the variability of night sky brightness leads to the identification of a fragmented clear sky, whereas the satellite data leads to a fully overcast night. From 2 a.m. the sky brightness was flat at around 22.0 mpsas or slightly above the average value though it was within the standard deviation. Indeed, the presence of clouds over unpolluted sites reduces the sky brightness, but the stable sky brightness and the drop in IR1 count lead to the expectation of a clear sky during that time.

The divergence between the SQM-based and satellite-based results hinders us from firmly confirming the ground truth. \emph{In-situ} observation usually is regarded as a better way to characterise an astronomical site compared to the remote sensing approach \citep{lagrosas2022}. However, zenithal sky brightness measurement is not an ultimate way of monitoring the sky, considering how narrow the field of view of the instrument is compared to the whole observable sky. In this regard, we consider the satellite-based results as the pessimistic solution, while the optimistic solution comes from the SQM data analyses. Table \ref{tab:results} summarises the values.

\begin{figure*}
    \centering
    \includegraphics[width=0.9\textwidth]{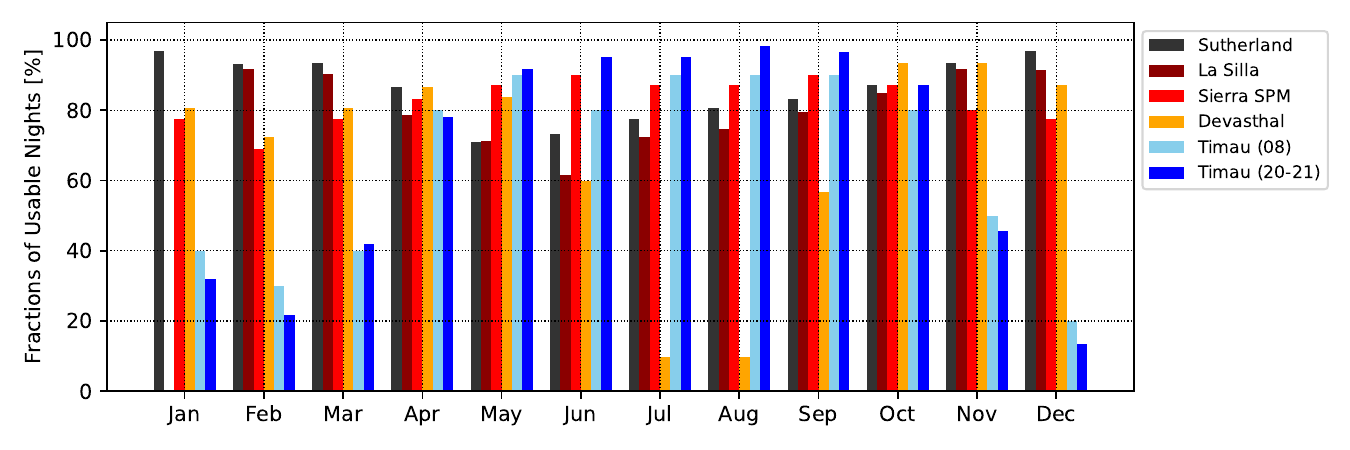}
    \caption{The percentages of usable nights at Timau and some other astronomical sites. Timau (20-21) is from the current study, Timau (08) data is from \citet{hidayat2012}, La Silla is from \citet{cavazzani2020}, while the Sutherland, Devasthal, and Sierra San Pedro Martir are from ERA5 reanalysis by \citet{ningombam2021}.}
    \label{fig:un_comparison}
\end{figure*}

The satellite-based results are in good agreement with 15-year statistics of clear sky fraction from \citet{hidayat2012}, with a mean absolute difference of $9\%$ (see also Fig. \ref{fig:un_comparison}). It is noteworthy that \citet{hidayat2012} analysed the clear sky fraction in spatial sense (e.g., $75\%$ of the night sky is clear at a certain night) while our study focuses on the temporal domain (e.g., $75\%$ of the night is clear). Nevertheless, both analyses lead to comparable percentages of usable nights both monthly and annually. Note also that a recent study using {\it FengYun-2} Satellite with a different methodology by \citet{wang2022} has led to consistent results with \citet{hidayat2012}.

Based on the satellite data, the expected percentage of clear sky over Timau is at least $66\%$ while the average AOT is $5.3$ hours per night. The average duration is prolonged to 7.9 hours during the dry season. It is approximately $87\%$ of the whole night between astronomical twilights. The lack of SQM data for February to April makes the annual average cannot be computed trivially. However, if we assume a constant value of usable nights for January to April, we get a yearly percentage of usable nights to be $74\%$. From May to October, the expected percentage of usable nights is $94\%$, among which $40\%$ are photometric nights with more than 6 hours of clear sky.

\section{Strategic values of the Timau site}
According to the general characteristics discussed above, we can consider the strategic values of the Timau site more thoroughly. This site is located on the West of the Pacific Ocean, while some major survey telescopes operate on the east side. Consequently, Timau is expected to gain the first opportunity to perform follow-up observations on some interesting transient objects discovered in some systematic surveys, such as Vera Rubin Telescope and Zwicky Transient Factory. Situated at a longitude of ${\sim}120^{\circ}$, Timau is also expected to fill the gap in the network of observatories that work for the International Asteroid Warning Network \citep[e.g.,][]{kofler2019, farnocchia2022}. Additionally, the Timau site provides relatively broad coverage of the night sky in the southern and northern hemispheres as it is located close to the equator. If we set a minimum altitude limit of $10^{\circ}$ above the horizon, we will get an observable sky covering more than 70\% of the whole sky. These geographical advantages are balanced with the fact that the tropical atmosphere tends to have more turbidity and more cloud cover \citep{aksaker2020}.

\begin{figure*}
    \centering
    \includegraphics[height=8cm]{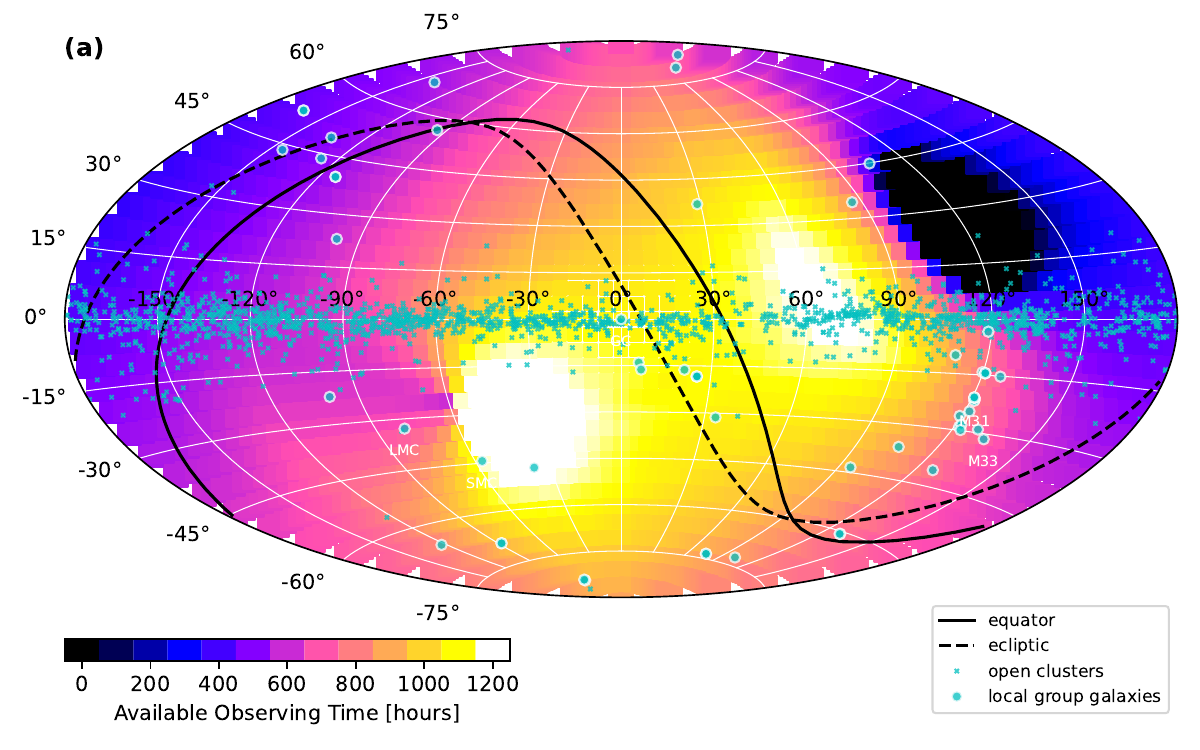}
    \includegraphics[height=8cm]{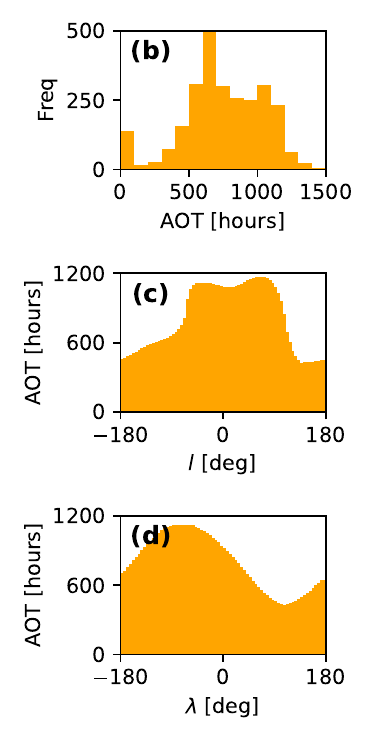}
    \caption{The observability map for the Timau shows the AOT of sky segments in Galactic coordinate (a). The location of open clusters \citep{dias2021} and local group galaxies are indicated by crosses and circles. The number distribution of sky segments as a function of AOT is also provided (b). The average AOT of sky segments as a function of galactic longitude ($l$) and ecliptic longitude ($\lambda$) are provided in panels (c) and (d). The average values are from the segments within $\pm30^{\circ}$ from the galactic equator or ecliptic.}
    \label{fig:observable}
\end{figure*}

A comparison between the percentages of usable nights at Timau and some other astronomical sites is presented in Fig. \ref{fig:un_comparison}. This plot is based on the data from \citet{cavazzani2020} for La Silla and \citet{ningombam2021} for the remaining sites. During the dry season (May-October), the Timau sky is almost transparent with the percentages of usable nights comparable to the Sierra San Pedro Martir in Mexico. Timau can be a complementary site to other sites such as Sutherland in South Africa and La Silla in Chile, where the fractions of usable nights slightly drop during winter. Timau can also complement the Devasthal in India, where a 3.6-metre optical telescope is installed \citep{sagar2019}. The percentages of usable nights at Devasthal drop to the minimum of ${\sim}20\%$ during the rainy season, which lasts from June to September.

Regionally, Timau is expected to complement some observatories where the observing seasons are around December. Thai National Observatory in Chiang Mai, Thailand, is mostly operated during the dry season from November to April \citep[e.g.,]{dhillon2014}. Similarly, the observing season for the Seimei Telescope at Okayama Astrophysical Observatory, Japan, is around the winter (October-March). In the southern hemisphere, however, the Siding Spring Observatory in Australia tends to have a clearer sky in April-September \citep{abbot2021}. This season coincides with the dry season at the Timau National Observatory.

Fig. \ref{fig:observable} displays a more detailed sky map from Timau with the AOT for each sky segment represented by colour. This map was created with the help of the \textsc{Astroplan Python} package \citep{morris2018}, which computed the observable time of a particular sky segment at a specific date. The altitude constraint was set to $10^{\circ}$ above the horizon, the time constraint was the astronomical twilights, and the minimum angular distance from the Moon was
\begin{equation}
\rho = 70q + 15\mu_{\text{lim}} - 285
\end{equation}
where $q$ represents the Moon’s fraction of illumination while $\mu_{\text{lim}}=18.0$ mpsas is the sky brightness limit. This simple formula approximates the Krischiunas-Schaefer model at the brightness level of $18.0$ mpsas (see Fig. \ref{fig:ks91}). Additionally, the monthly average of the AOT derived from the satellite data (see Table \ref{tab:results}) was used as the weighting factor to get a more realistic prediction of the available observing time for each sky segment.

For the whole sky, the median AOT is 760 hours/year, while the first and third quantiles are 580 and 940 hours/year. From the map, we can see that the observing season at Timau is apt for observing the Galactic centre though the lunar cycle slightly affects the observability. As indicated in the panel (c) of Fig. \ref{fig:observable}, the maximum chance of observing objects located between $-60^{\circ}$ to $90^{\circ}$ galactic longitude is about 1000 hours/year or more. Some local galaxies like the Small and Large Magellanic Cloud are located in this direction. The chances drop to less than 300 hours/year for the regions around the Anti-Galactic Centre.

In terms of sky brightness, Timau is considerably pristine with sky brightness $\mu_0\approx22.0$ mpsas. However, the threats of light pollution also come from the growing population of resident space objects (RSO), primarily due to the launch of satellite constellations \citep{green2022}. Many satellites in low Earth orbits are visible to the naked eye for a few hours after sunset and before sunrise. Those objects are real interferences for optical and infrared observations, either wide-field high-sensitivity imaging \citep{walker2020, hainaut2020} or low-angular-resolution observations \citep{kocifaj2021}. Astronomical sites at the middle latitudes (${\sim}50^{\circ}$ north and south) are the most impacted as the number of satellites above the horizon at any moment may reach 500 \citep{lawler2021, osborn2022}. Even though the equatorial region is not free from threats, the appearance of satellites over this region is reduced to ${\sim}80\%$ \citep{osborn2022}. From another perspective, Timau can also be a good site for RSO and space debris observation and characterisation \citep{silha2020, danarianto2022}.

Regardless of the fact that excited OH molecules contribute to the natural sky brightness and affect astronomical observations, there is also an opportunity for space weather research through systematic OH airglow monitoring. A study on Mesospheric gravity waves over a tropical convective region \citep{Nakamura2003, moral2019} can be a good example. Timau National Observatory was planned as a multi-purpose observatory where space weather research and monitoring are also part of its activities \citep{mumpuni2018}. Thus, OH airglow observation can be one of the strategies that can be contributed to international networks \citep[e.g.,][]{Li2020}.

\section{Conclusions}
In this study, we analyse \emph{in-situ} sky brightness measurements and remote sensing data from {\himawari} to characterise the Timau National Observatory in Indonesia. The main conclusions of this study are:
\begin{enumerate}
    \item Timau site has a pristine sky with a zenithal brightness of $\mu_0=21.86\pm0.38$ mpsas. At this moment, the threat from the neighbouring municipalities is minimal though preserving this condition in the future is mandatory.
    \item Based on the analysis of the moonlit sky brightness, we estimate the extinction coefficient $k=0.48\pm0.04$, which is relatively high compared to the figures from other observatories. However, this estimate is justifiable because Timau is situated in the low-altitude tropical region. The moonlit sky brightness also indicates that the aerosol content, which is responsible for Mie scattering, is elevated.
    \item Approximately 240 days ($66\%$) are available for observation throughout the year. The observing season lasts from May to October. During that period, the available observation time (AOT) is around $7.8$ hours per night.
    \item Additionally, the observing season at the Timau site is keen on the direction of the Galactic Centre. For the regions between $-60^{\circ}$ to $90^{\circ}$ Galactic longitude, the total AOT reaches 1000 hours per year.
    \item Timau National Observatory with its pristine situation is also provide opportunity for various research topics, including RSO and space debris observations, as well as sky glow dynamic for space weather study.
\end{enumerate}

Compared to previous studies \citep{hidayat2012,wang2022}, this study provides more detailed pictures of the site characteristics, which is helpful for strategic planning in the future.

\section*{Acknowledgment}
We acknowledge valuable comments from the referees. We thank Kochi University for providing processed {\himawari} images used in this study.\\
RP conceived the idea and performed most of the analyses and writing, MBS and MDM acquired and processed the \emph{in-situ} sky brightness data. All authors contributed to the final version of the manuscript.

\section*{Data Availability}
The data underlying this article are available in the \href{https://data.lipi.go.id/dataset.xhtml?persistentId=hdl:20.500.12690/RIN/A5XCJB}{RIN Dataverse}.

\bibliographystyle{mnras}
\bibliography{main}

\label{lastpage}
\end{document}